\documentclass[pdflatex,sn-mathphys-num]{sn-jnl}

\usepackage{graphicx}%
\usepackage{multirow}%
\usepackage{amsmath,amssymb,amsfonts}%
\usepackage{amsthm}%
\usepackage{mathrsfs}%
\usepackage[title]{appendix}%
\usepackage{xcolor}%
\usepackage{textcomp}%
\usepackage{manyfoot}%
\usepackage{booktabs}%
\usepackage{algorithm}%
\usepackage{algorithmicx}%
\usepackage{algpseudocode}%
\usepackage{listings}%

\theoremstyle{thmstyleone}%
\theoremstyle{thmstyletwo}%

\theoremstyle{thmstylethree}%

\begin{document}

\title{Polarization-Resolved Chlorophyll Imaging for Non-Invasive Plant Tissue Assessment Using a Silicon-Rich Nitride Metalens Array}


\author[1]{\fnm{Alireza} \sur{Khalilian}}\email{akhalili@umich.edu}

\author[3]{\fnm{Mehdi} \sur{Sh. Yeganeh}}\email{mehdiyeg@umich.edu}

\author[1]{\fnm{Konstantin} \sur{Tsoi}}\email{kostyat@umich.edu }

\author[1]{\fnm{Bowen} \sur{Yu}}\email{bowenyu@umich.edu }

\author[3]{\fnm{Joe} \sur{Lo}}\email{jfjlo@umich.edu }

\author*[1,2]{\fnm{Yasha} \sur{Yi}}\email{yashayi@umich.edu}

\affil*[1]{\orgdiv{Integrated Nano Optoelectronics Laboratory}, \orgname{University of Michigan-Dearborn}, \orgaddress{\street{4901 Evergreen Rd.}, \city{Dearborn}, \postcode{48128}, \state{MI}, \country{USA}}}

\affil[2]{\orgdiv{Energy Institute, University of Michigan}, \orgname{Organization}, \orgaddress{\street{2301 Bonisteel Blvd.}, \city{Ann Arbor}, \postcode{48109-2100}, \state{MI}, \country{USA}}}

\affil[3]{\orgdiv{Department of Mechanical Engineering}, \orgname{University of Michigan-Dearborn}, \orgaddress{\street{4901 Evergreen Rd.}, \city{Dearborn}, \postcode{48128}, \state{MI}, \country{USA}}}

\abstract{Polarization-sensitive imaging enhances contrast and reveals structural features in biological tissues that are often missed by intensity-based methods, but its adoption is limited by bulky optics. We present a compact silicon-rich nitride (SRN) metalens array for high-resolution, polarization-resolved imaging of plant tissue at the chlorophyll absorption peak (660\,nm). The array integrates orthogonally sensitive metalenses to simultaneously capture X- and Y-linearly polarized transmission images, enabling real-time, label-free assessment of plant microstructure and stress responses. Polarization fusion and difference mapping reveal structural anisotropy and pigment variation in both healthy and stressed leaves. The SRN metalens, designed via an inverse approach using birefringent meta-atoms and fabricated through CMOS-compatible processes, achieves a large numerical aperture, high transmission, and spectral alignment with biological absorbers. This work demonstrates the feasibility of compact, integrated polarization-resolved imaging, offering a scalable alternative to conventional systems. The approach holds potential for biomedical and agricultural applications, where detecting subtle polarization-dependent changes could enable early diagnosis and tissue characterization.}




\maketitle

\section*{Introduction}

Imaging biological tissue is crucial for identifying early indicators of disease and understanding the progression of physiological disorders. Optical imaging is particularly valuable due to its ability to provide internal insights without causing physical harm. As a non-invasive approach, it enables real-time monitoring in diverse biological and clinical contexts. Over the years, a range of optical imaging techniques has been developed to address this need, including fluorescence microscopy, confocal microscopy, and hyperspectral imaging \cite{xu2024multiphoton, lai2024advancing}. Among current methods, polarization-sensitive imaging is notable for its ability to reveal structural and functional attributes frequently missed by conventional intensity- or spectrum-based techniques \cite{tang2020polarization}. Subtle characteristics such as tissue anisotropy, fiber orientation, and stress-induced modifications can be elucidated by examining the impact of polarized light on tissue, encompassing birefringence, depolarization, and dichroism \cite{suzuki2025acquisition}. However, relying solely on intensity or spectral imagery often fail to adequately disclose these attributes \cite{oldenbourg1996new, jermain2024deep}. This feature is especially effective for assessing plant health, offering a biologically relevant but experimentally accessible framework to illustrate the benefits of polarization-resolved imaging. Plants exhibit characteristic polarization responses arising from their internal microstructure and natural birefringence, which can change in magnitude or pattern under stress or nutritional deficiencies \cite{rodriguez2022polarimetric, li2022new}. Such conditions often alter chloroplast organization, cell wall structure, and pigment content, all of which influence both polarization signatures and photosynthetic performance \cite{van2021polarimetric}. Chlorophyll, the primary pigment driving photosynthesis, strongly absorbs light near $\lambda = 660\,\mathrm{nm}$ and $\lambda = 680\,\mathrm{nm}$, making these wavelengths particularly suitable for assessing its concentration and spatial distribution \cite{lichtenthaler2001chlorophylls}. While tools such as SPAD meters and spectrophotometers are frequently used to evaluate chlorophyll content, they have notable limitations. A SPAD (Soil Plant Analysis Development) meter is a handheld device that estimates relative chlorophyll concentration by measuring leaf transmittance at two wavelengths, but it is confined to small, localized measurement areas. Spectrophotometry, in contrast, often requires destructive sample preparation \cite{wicharuck2024implementation, chazaux2022precise}. Hyperspectral imaging systems can provide extensive spectral information; however, their size, cost, and operational complexity often restrict their use to controlled laboratory environments \cite{zolotukhina2023extraction}. In contrast, polarization-resolved imaging facilitates label-free, spatially resolved evaluations of chlorophyll and tissue structure through the use of compact setups. Even simplified dual-polarization approaches—relying on orthogonal linear states—can uncover significant distinctions through differential analysis methods, such as absolute or signed difference mapping \cite{rodriguez2022polarimetric}. The methods emphasize characteristics that traditional techniques might miss, enhancing the contrast between stressed and healthy areas and facilitating subsequent image processing by offering insights into tissue orientation and structural integrity \cite{van2021polarimetric}. 

To improve the accessibility and scalability of polarization-resolved imaging, recent research efforts have focused on metasurface optics. Metalenses—ultrathin optical components made from carefully designed subwavelength nanostructures—offer precise control over the phase, amplitude, and polarization of light, all within a compact and lightweight form factor \cite{li2025flat, chen20253d, fu2025miniaturized}. Their planar geometry and compatibility with common lithographic fabrication techniques render them especially suitable for integration into portable imaging devices and CMOS-based platforms. Nonetheless, the broader adoption of metalenses has been hindered by material and fabrication limitations. Commonly used materials— such as titanium dioxide (TiO$_2$) \cite{li2025dielectric}, amorphous silicon (a-Si) \cite{zhang2022high}, and gallium nitride (GaN) \cite{chen2021high}, often require fabrication steps that are either too complex or not compatible with standard CMOS processes, making them less suitable for large-scale production. Silicon-rich nitride (SRN) has recently drawn interest as a practical material for metalens fabrication. It offers a high refractive index, minimal optical loss in the visible range, and strong compatibility with standard CMOS processes \cite{goldberg2024silicon, khalilian2025chlorophyll}. These features make SRN a compelling choice for creating metalenses that are both efficient and easier to scale—qualities that are important for transitioning from lab-scale demonstrations to real-world applications in research and industry \cite{froch2025beating, ye2019silicon}.

This study presents a high-numerical-aperture (NA) metalens array fabricated using silicon-rich nitride (SRN), carefully designed for dual-polarization imaging utilizing orthogonal linear polarization states (XLP and YLP). We demonstrate its potential via chlorophyll absorption imaging, serving as a proof of concept for non-invasive evaluation of plant tissue characteristics. By combining SRN’s CMOS-compatible fabrication with the structural sensitivity of polarization imaging, the resulting platform offers a lightweight, scalable, and biologically relevant solution with broad applicability in precision agriculture, environmental monitoring, and biomedical diagnostics.

\section{Metalens Design Methodology}\label{sec2}

\begin{figure}[H]
\centering\includegraphics[width=13cm]{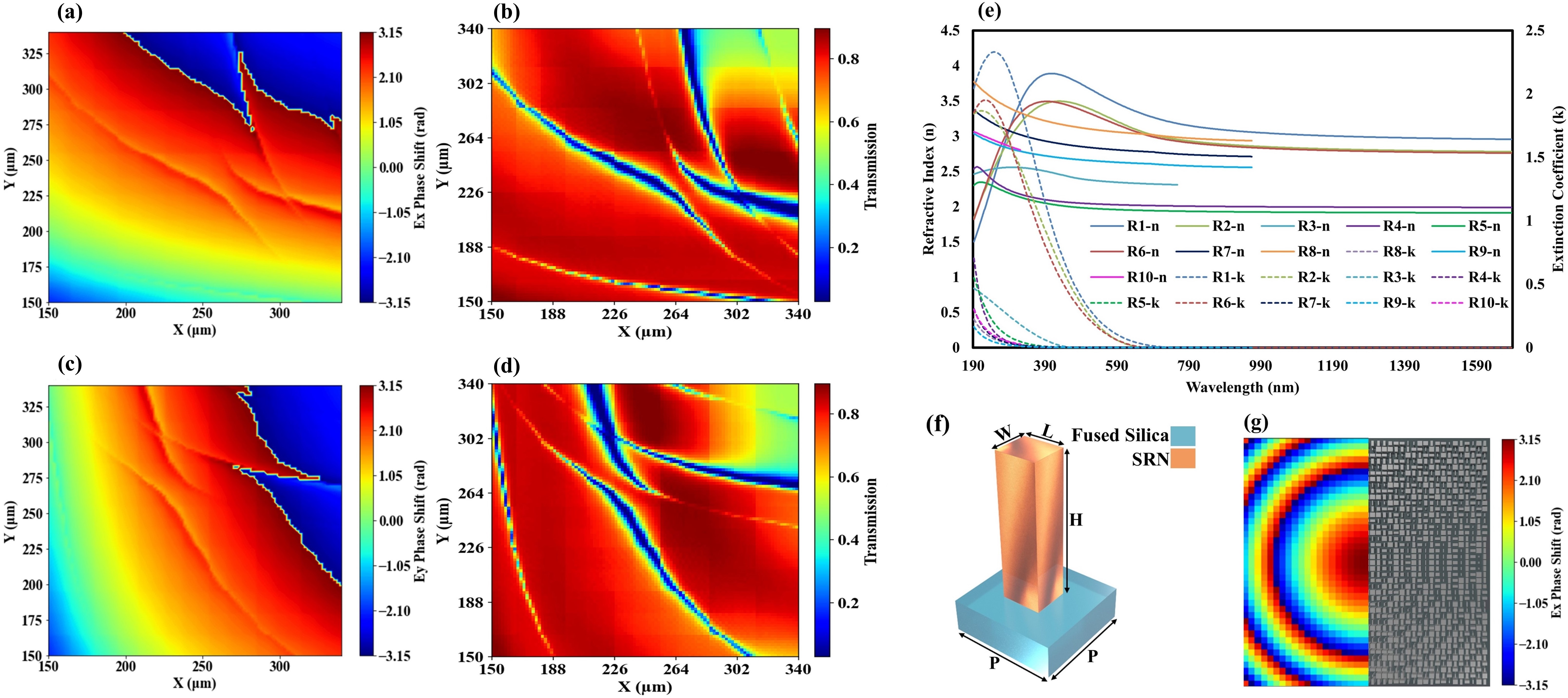}
\caption{(a, c) Simulated phase shift maps for XLP and YLP, and (b, d) transmission maps for the XLP and YLP lens, respectively. (e) Measured refractive index ($n$) and extinction coefficient ($k$) of SRN films for different deposition recipes (R denotes recipe). (f) Schematic of a meta-atom. (g) Target phase profile and 3D lens layout.}
\label{fig:simulation}
\end{figure}

A high-numerical-aperture (NA) metalens was designed to operate at a wavelength of $\lambda = 660\,\mathrm{nm}$. The focusing function was achieved by implementing a spatially varying phase mask, where each point on the lens surface introduces a specific phase delay to shape the transmitted wavefront into a diffraction-limited focus. The target phase distribution follows a hyperboloidal profile (Figure~\ref{fig:simulation}(g)),
\begin{equation}
    \Phi(x,y) = \frac{2\pi}{\lambda}\!\left(f - \sqrt{f^2 + x^2 + y^2}\right),
\end{equation}
which ensures constructive interference at the focal plane. With $\mathrm{NA} = 0.5$, the calculated focal length is $f \approx 52.7\,\mu\mathrm{m}$.

The lens aperture was divided into a square grid containing $156 \times 156$ meta-atoms, each separated by a pitch of $390\,\mathrm{nm}$, giving an effective aperture diameter of about $60.8\,\mu\mathrm{m}$. The meta-atom geometries were parameterized by their lateral dimensions - length ($l$) and width ($w$) with fixed height (H) of $600\,\mathrm{nm}$ (Figure~\ref{fig:simulation}(f)) - and selected from a simulation library generated using Lumerical FDTD. This library was obtained by sweeping $l$ and $w$ from $150\,\mathrm{nm}$ to $340\,\mathrm{nm}$, producing $100$ samples along each dimension. For every geometry, the simulated phase shift $\phi(l,w)$ and transmission efficiency $T(l,w)$ were recorded for both $x$-linearly polarized (XLP) (Figure~\ref{fig:simulation}(a) and (b), respectively) and $y$-linearly polarized (YLP) (Figure~\ref{fig:simulation}(c) and (d), respectively) light.\noindent The primary rationale for selecting silicon-rich nitride (SRN) as the lens material stems from our recent findings that high-index dielectrics permit smaller unit-cell pitch, which directly improves metalens focusing efficiency \cite{khalilian2025high}. Material optical properties of the silicon-rich nitride (SRN) layer were measured experimentally at $660\,\mathrm{nm}$ using a Woollam M-2000 spectroscopic ellipsometer. The deposition process parameters—such as gas flow ratios, chamber temperature, and process pressure were tuned to achieve a high refractive index with minimal absorption. The chosen recipe provided $n = 2.71$ and $k = 0.005$ at the design wavelength. Compared with widely used dielectrics such as \(\mathrm{TiO_2}\) and GaN at this regime, this combination of high index and low extinction coefficient is advantageous for high-performance metalens design.  Optical constants for several deposition recipes are shown in Figure~\ref{fig:simulation}(e). The FDTD simulations used a uniform spatial mesh of $15\,\mathrm{nm}$ in all three dimensions to ensure accurate representation of the nanostructures. Additional details of the simulation procedures for both unit-cell and full-device models are available in \cite{khalilian2025high}. The simulated phase data were unwrapped to remove discontinuities and normalized into the range $[0, 2\pi)$:

\begin{equation}
    \phi_{\mathrm{norm}}(l,w) = \bmod\!\left(\phi(l,w) - \min\{\phi(l,w)\}, 2\pi\right)
\end{equation}

Cubic interpolation was then applied to allow continuous sampling across the $(l,w)$ design space. To minimize propagation losses, only geometries with $T(l,w) \ge 0.5$ were considered for the final design. An inverse-design process was applied across the lens surface. For each lattice position $(x,y)$, the meta-atom dimensions $(l^*,w^*)$ were chosen to minimize the local phase mismatch with respect to the target phase profile:
\begin{equation}
    (l^*, w^*) = \arg\!\min_{(l,w)} \left|\phi(l,w) - \Phi(x,y)\right|,
    \quad \text{subject to } T(l,w) \ge 0.5.
\end{equation}
The modulo operation ensured phase continuity within $[0, 2\pi)$. A circular aperture of radius $R \approx 30.4\,\mu\mathrm{m}$ was imposed by retaining only points where $\sqrt{x^2 + y^2} \le R$. The selected geometries were compiled into a GDSII layout file using automated scripts for fabrication by electron-beam lithography.

\section*{Fabrication}\label{sec3}

\begin{figure}[htbp]
\centering
\includegraphics[width=13cm]{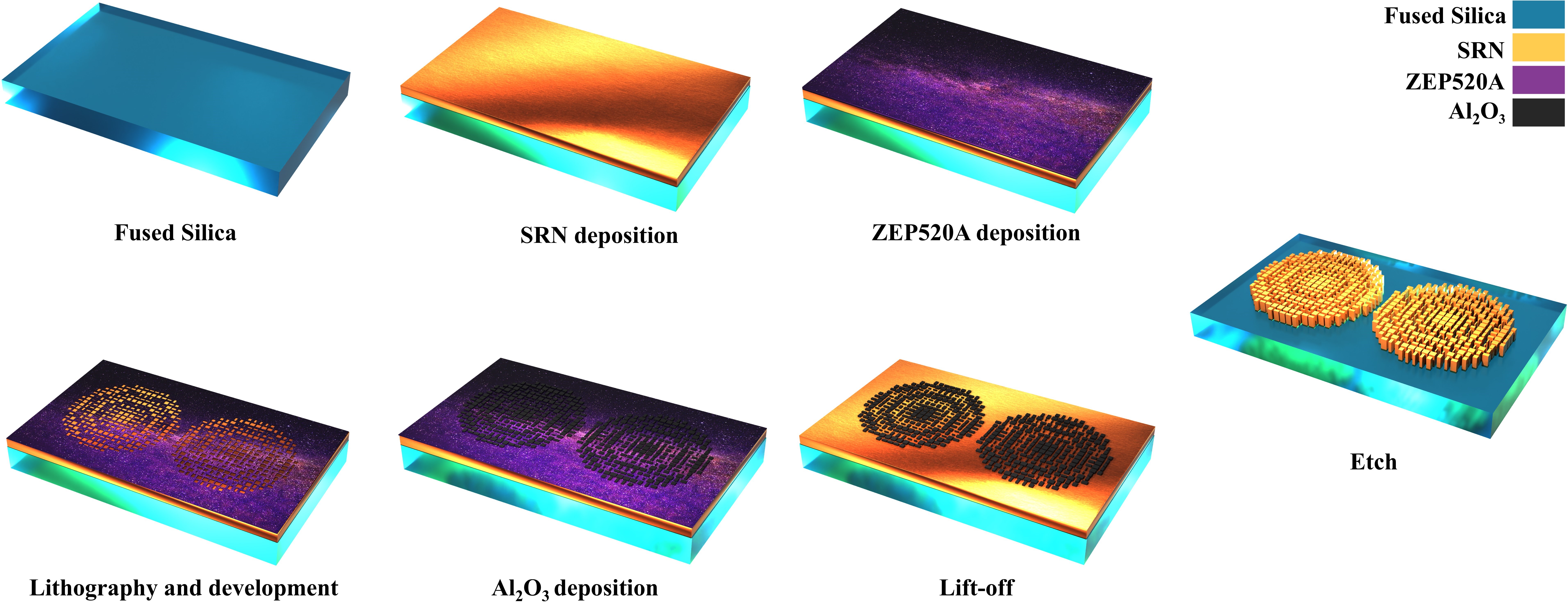}
\caption{Schematic overview of the metalens fabrication process.}
\label{fig:fabrication process}
\end{figure}

Double-sided polished 4-inch fused silica wafers with a nominal thickness of 0.5 mm were purchased from SVM. The precise thickness of each wafer was quantified to provide accurate input for curve fitting in the ellipsometry process.  Preliminary cleaning was conducted with isopropyl alcohol (IPA), followed by an oxygen plasma treatment in a YES Plasma Stripper utilizing the Descom process for about 15 seconds to remove organic impurities and enhance surface adhesion during the PECVD process. The comprehensive fabrication process flow is shown in Figure~\ref{fig:fabrication process}.

A 600 nm-thick layer was then deposited on the fused silica wafer using a custom recipe on the GSI ULTRADEP 2000 PECVD. The inherent stress of the deposited films across four different was ~60 MPa (measured  with Dektak XT profilometer) showing no cracks or wafer bowing. The wafers were afterwards cut into nearly 1-inch by 1-inch pieces. Each sample underwent optical property characterization and thickness verification of the SRN layer using a Woollam M-2000 spectroscopic ellipsometer, verifying agreement with design simulations and providing accurate parameters for the etching process. Before applying the resist layer, the samples underwent an additional 15-second oxygen plasma treatment in the YES Plasma Stripper. A diluted ZEP520A (1:1) solution was spin-coated at 1000 rpm for 45 seconds to obtain a resist thickness of approximately 300 nm, and was thereafter baked at 180°C for 2 minutes. To reduce the charging effects caused by transparent substrates, DisCharge H$_2$O was spin-coated at 4500 rpm for 45 seconds. Electron-beam lithography was performed using a dose of 240~$\mu$C/cm$^2$, a beam current of 500~pA, and the fifth lens mode. Following exposure, samples were rinsed with deionized water to eliminate the DisCharge H$_2$O layer, then immersed in a cold ZED-N50 bath (about 0$^\circ$C) for 30 seconds, followed by a 30-second immersion in IPA bath and a comprehensive IPA rinsing. The resist thickness was confirmed post-development using a KLA profilometer. An oxygen plasma cleaning for 15 seconds was performed to remove the remaining resist (YES Plasma Stripper, Descom recipe). An aluminum oxide (Al$_2$O$_3$) hard mask, about 60 nm in thickness, was subsequently deposited via electron-beam evaporation (Angstrom Engineering Evovac) at a chamber pressure of roughly 3.96$\times$10$^{-6}$~Torr and a deposition rate of 0.5~\AA/s. Lift-off was executed by submerging the samples in a Remover PG bath for 24 hours, then subjected to sonication at 80°C for 2 hours in fresh Remover PG, and rinsed with IPA. Pattern transfer into the SRN layer was accomplished utilizing an STS APS DGRIE system (STS Glass Etcher), an ICP-RIE apparatus operated with a gas mixture of SF$_6$ and C$_4$F$_8$. Post-etch inspections verified that negligible Al$_2$O$_3$ residues remained on the surface. Figure~\ref{fig:fabrication process}(a) shows the fabricated metalens next to a coin for scale reference. The optical microscope images of the lens array (Figure~\ref{fig:fabrication process}(b)) highlight the polarization-sensitive lenses, with red and blue circles representing XLP- and YLP-sensitive lenses, respectively. Scanning electron microscopy (SEM) images in Figure~\ref{fig:fabrication process}(c-e) reveal the fabricated meta-atom arrangement of the metalens at various magnifications.

\begin{figure}[H]
\centering\includegraphics[width=13cm]{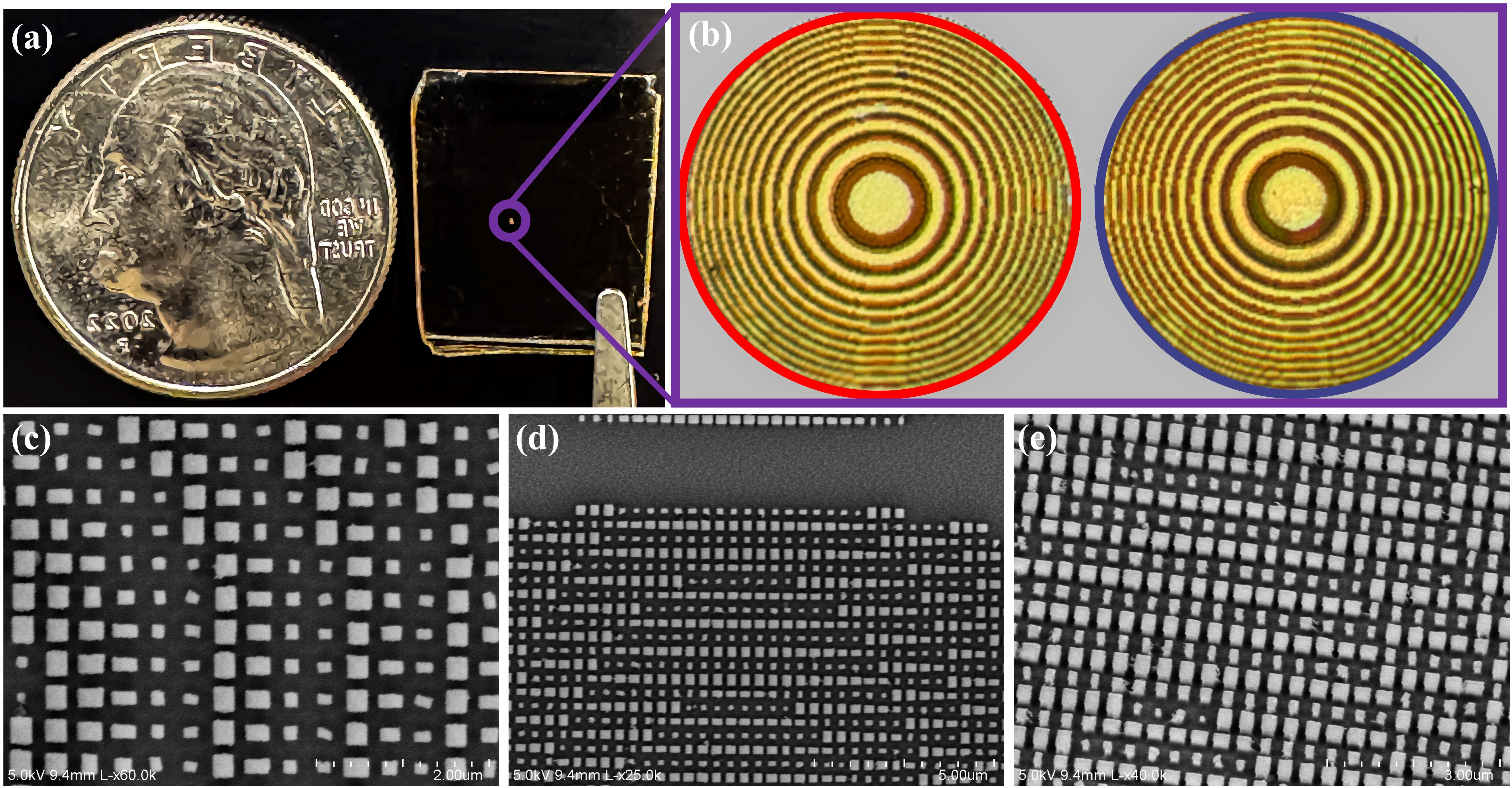}
\caption{(a) Fabricated metalens chip beside a coin for scale. (b) Optical images of the fabricated lens array: red circle denotes XLP-sensitive lens, blue circle denotes YLP-sensitive lens. (c–e) SEM images of the fabricated metasurface showing meta-atom arrangement at different magnifications.}
\label{fig:fabrication sem}
\end{figure}

\section{Results and Discussion}

\subsection{Metalens Characterization}

\begin{figure}[H]
\centering
\includegraphics[width=13cm]{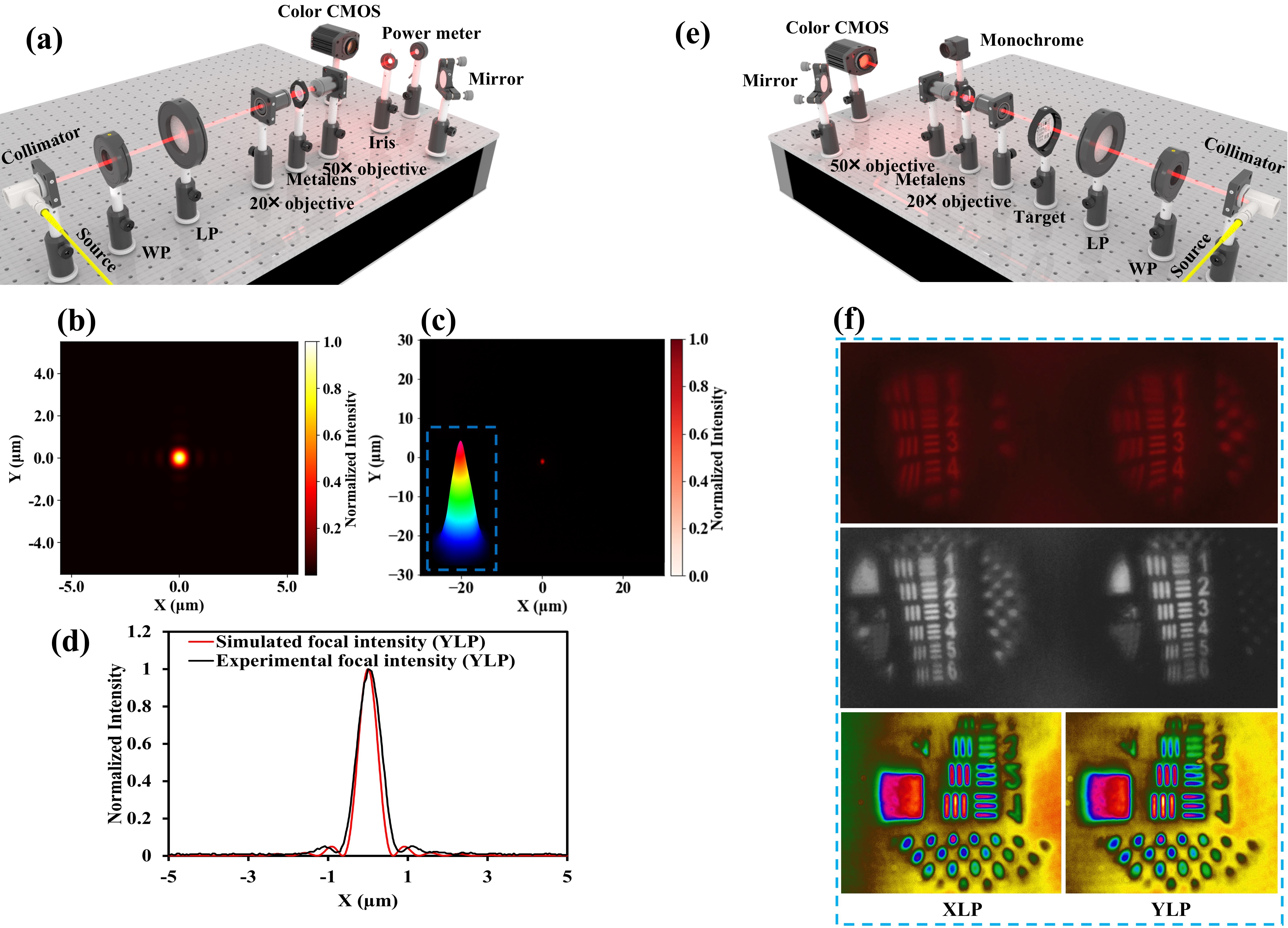}
\caption{Polarization-selective metalens array at 660 nm. (a,e) Experimental layouts for focus and target imaging. (b) Simulated YLP focal intensity. (c) Measured focal spot at the design plane using a dual scanning slit beam profiler (BP209-VIS); color banding is false-color rendering. (d) Normalized line profile: simulation (red) vs. experiment (black). (f) Simultaneous XLP and YLP views of a USAF resolution target—top: color-CMOS frames; middle: raw BMP frames from the high-resolution monochrome sensor used for all lateral analyses; bottom: pseudo-colored renderings highlighting polarization-dependent sensitivity.}
\label{fig:results1}
\end{figure}

A comprehensive characterization of the dual-channel polarization-selective metalens array at 660~nm was performed using the configurations in Figure~\ref{fig:results1}(a) and (e). Focal metrics were first measured without a target. Figure~\ref{fig:results1}(b, c) are showing simulated YLP focal intensity and the measured focal spot captured with the monochrome sensor and the through-focus beam caustic captured using a dual scanning slit beam profiler (BP209-VIS). For both XLP and YLP, the experimental full width at half maximum (FWHM) was $\sim 1.06\lambda$, close to the simulated $0.89\lambda$ for a 10~$\mu$m lens (Figure~\ref{fig:results1}(d)). The measured focal length for both channels was $46.7~\mu\mathrm{m}$, corresponding to an experimental numerical aperture of $0.54$, in reasonable agreement with the designed value of $0.5$. Focusing efficiency---calculated as the ratio between the optical power measured in the focal plane within a circular aperture of diameter $3\times\mathrm{FWHM}$ and the total incident optical power of the lens, both measured using a Thorlabs PM100D power meter with an S120C photodiode sensor--was $65.9\%$ for YLP and $55.8\%$ for XLP, compared to $75\%$ in simulation. Imaging performance was then assessed with a USAF resolution target per Figure \ref{fig:results1}(e). XLP and YLP images are shown in Figure \ref{fig:results1}(f): top row, color-CMOS frames for visualization; middle row, raw BMP frames from the high-resolution monochrome sensor used for all lateral analyses; bottom row, pseudo-colored renderings that highlight polarization-dependent sensitivity while preserving co-registration. In the Figure \ref{fig:results1}(f), the normalized pseudo-colored intensity maps in the bottom row reveal orientation-dependent polarization response, where horizontal bar groups appear brighter in the XLP channel, whereas vertical bar groups appear brighter in the YLP channel. This behavior is consistent with polarization-dependent diffraction efficiency, where features oriented perpendicular to the incident electric-field vector exhibit enhanced transmission. In the XLP channel, horizontal features show stronger core brightness and sharper edges, while in the YLP channel, vertical features display higher perceived intensity and more uniform filling across the bars. These complementary responses illustrate the capability of the dual-channel metalens array to capture co-registered, polarization-resolved datasets, enabling improved feature discrimination and enhanced orientation-specific contrast that would otherwise be masked in unpolarized imaging.

\subsection{Polarization-resolved intensity analysis}

Polarization-resolved imaging in this work can be regarded as a reduced implementation of Mueller-matrix polarimetry, in which the sample’s polarization-dependent transmission is assessed through its influence on selected Stokes parameters. A general polarization state is described by the Stokes vector $(S_0,S_1,S_2,S_3)$, whereas a full Mueller-matrix measurement characterizes how these quantities transform upon interaction with the specimen. Here, we adopt a dual-linear acquisition scheme that records $S_0$ (total intensity) and $S_1$ (difference between horizontal and vertical components), following prior reports that have demonstrated its effectiveness in revealing diagnostically relevant birefringence and dichroism in biological tissues~\cite{song2021stokes, qi2017mueller}. This approach retains the essential sensitivity to anisotropy and orientation while avoiding the complexity and acquisition time required for full-Stokes or Mueller protocols~\cite{van2019depolarizing}.

\subsubsection{Image-processing pipeline}
To enable pixel-wise comparison, XLP and YLP images were first registered using subpixel enhanced-correlation-coefficient (ECC) alignment and cropped to their common field to remove background and edge artifacts. Let \(I_x(x,y)\) and \(I_y(x,y)\) denote the matched XLP and YLP frames; from these, we computed three derived maps for both the conventional objective-lens and metalens images.

\begin{equation}
I_{\mathrm{Unpol Avg}}(x,y)=\frac{I_x(x,y)+I_y(x,y)}{2},
\label{eq:Iunpol}
\end{equation}
\smallskip
\begin{equation}
\mathrm{AbsDiff}(x,y)=\left|I_y(x,y)-I_x(x,y)\right|,
\label{eq:Abs Diff}
\end{equation}
\smallskip
\begin{equation}
\mathrm{SignedDiff}(x,y)=I_y(x,y)-I_x(x,y),
\label{eq:SignedDiff}
\end{equation}

\smallskip

For ideal orthogonal linear analyzers,
\(I_x=(S_0+S_1)/2\) and \(I_y=(S_0-S_1)/2\). Consequently,
eq.~\eqref{eq:Iunpol} estimates the unpolarized intensity (proportional to \(S_0\)); we report the
mean rather than the sum to keep the dynamic range comparable to a single channel.
Moreover, eq.~\eqref{eq:Abs Diff} is proportional to \(|S_1|\), highlighting polarization-sensitive
structure irrespective of polarity, while eq.~\eqref{eq:SignedDiff} is proportional to \(-S_1\)
(the sign depends on which channel is labeled \(x\) vs. \(y\)). Edge maps were extracted with the \(3\times3\) Sobel operator \cite{kanopoulos1988design}.
For each input image \(J\in\{I_x,I_y,I_{\text{unpol}}\}\), we first min--max normalized \(J\) to \([0,1]\)
and computed a single directional Sobel derivative (column direction, reflect padding).
The resulting response was used as the edge-strength image for visualization; intensity
ranges were clipped at a fixed percentile for display consistency across datasets.

\subsubsection{Conventional objective-lens validation}

\begin{figure}[H]
\centering
\includegraphics[width=13cm]{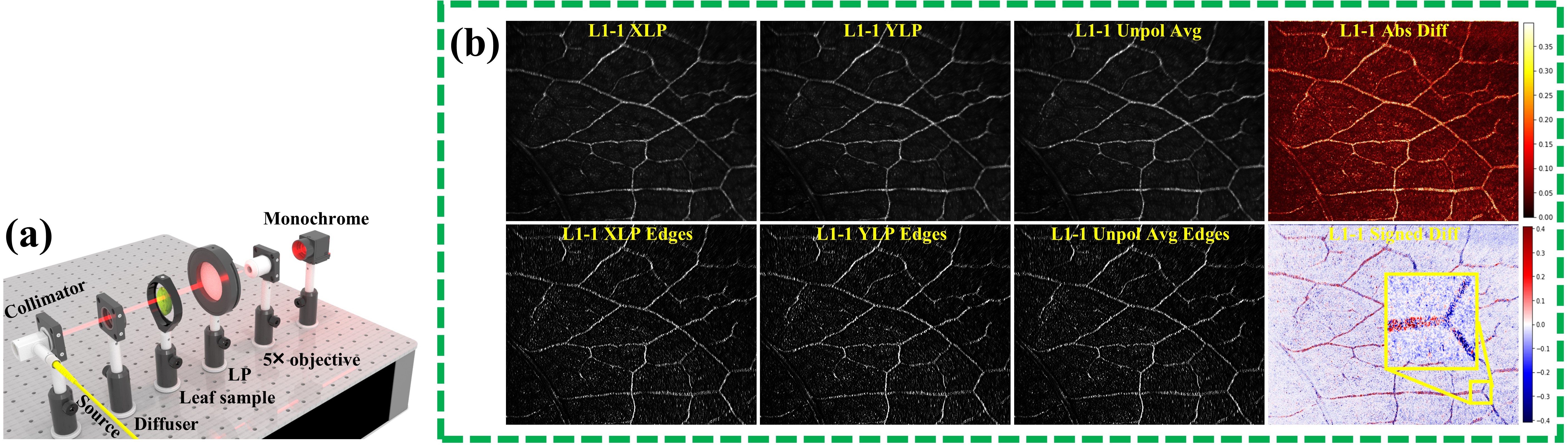}
\caption{Polarization-resolved imaging with a 5× objective lens. (a) Experimental setup. (b) Processed images of leaf region L1-1 showing XLP, YLP, Unpol Avg, Abs Diff, Signed Diff, and edge maps, highlighting enhanced vein contrast and structural details.}
\label{fig:objective results}
\end{figure}

We first validated the effectiveness of polarization sensitivity using a conventional transmission microscope (Figure~\ref{fig:objective results}(a)). The setup employed collimated $660\,\mathrm{nm}$ illumination with a diffuser to suppress speckle, a $5\times$ objective, a rotatable linear polarizer (analyzer), and a monochrome camera. We denote \emph{leaf samples} by $L_k$ ($k=1,2,\ldots$) and fields of view by $L_k$--$m$; thus, $L1$--$1$ refers to sample 1, field of view 1. From the same field of view of leaf sample $L1$ shown in Figure~\ref{fig:results2}(c), two polarization-resolved images were recorded with the analyzer oriented at $0^\circ$ (XLP) and $90^\circ$ (YLP), shown in Figure~\ref{fig:objective results}(b) as $L1$--$1$ (XLP) and $L1$--$1$ (YLP), respectively. Distinct vein segments and aligned fiber bundles exhibited pronounced intensity modulation: features parallel to the analyzer axis appeared bright, whereas orthogonal structures were attenuated. Several features reversed contrast between the two channels, consistent with linear birefringence and possible contributions from linear dichroism and orientation-dependent scattering in the cellulose microfibril network. These results demonstrate that polarization-resolved imaging can reveal orientation-specific details and enhance structural contrast relative to unpolarized averaging, thereby motivating the replacement of conventional bulk optics with an integrated metalens system that delivers simultaneous, co-registered polarization-resolved imaging in a compact form factor.

\subsubsection{Integrated metalens acquisition}

\begin{figure}[H]
\centering
\includegraphics[width=10cm]{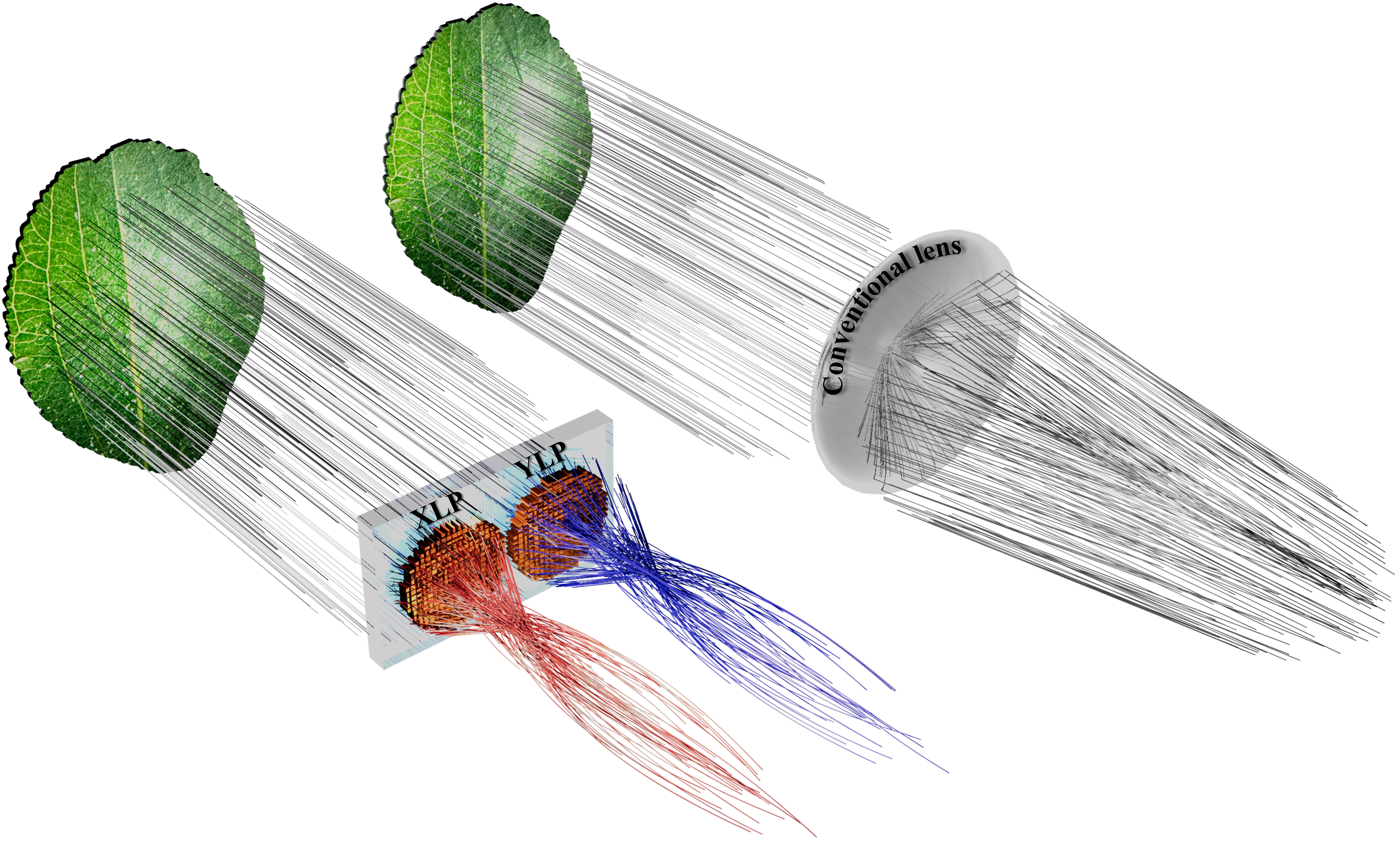}
\caption{3D illustration of the comparison between a conventional lens and a polarization-sensitive metalens array for plant tissue imaging. }
\label{fig:array vs conv}
\end{figure}

\begin{figure}[H]
\centering
\includegraphics[width=13cm]{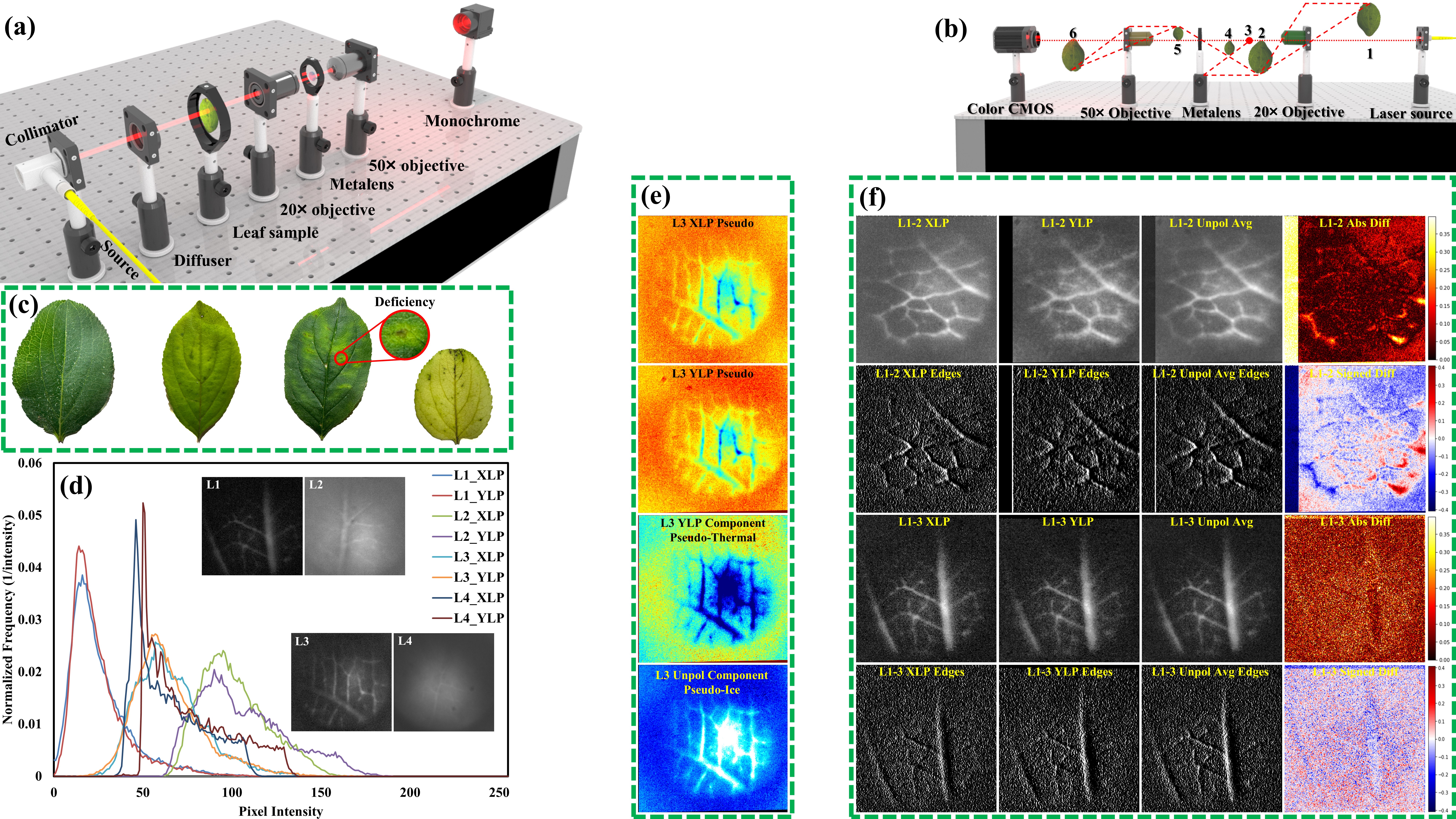}
\caption{Polarization-resolved imaging of Prunus cerasifera leaves using the SRN metalens array. (a) Metalens imaging setup with dual XLP–YLP channels. (b) Ray diagram showing optical planes 1–6 from source to imaging sensor. (c) Leaf samples L1–L4 representing healthy, moderately stressed, early pigment-deficient, and severely chlorophyll-depleted conditions. (d) Normalized pixel intensity histograms for XLP and YLP with representative images. (e) Pseudo-colored visualization of polarization-sensitive and polarization-suppressed components for L3, showing early pigment deficiency. (f) Processed XLP, YLP, Unpol Avg, Abs Diff, Signed Diff, and Edge maps for subregions L1-2 and L1-3 of the same healthy leaf sample.}
\label{fig:results2}
\end{figure}

Having established the contrast mechanism, we employed the fabricated SRN metalens array, which integrates focusing and polarization analysis into a single optical element. Figure~\ref{fig:array vs conv} shows a side to side functionality of the proposed polarization sensitive metalens array and a conventional lens. The conventional lens collects and combines all polarization components into a single image, potentially averaging out orientation-dependent structural information. However, the polarization-sensitive metalens array separates XLP and YLP components, preserving polarization-resolved information and enabling enhanced structural contrast and orientation-specific analysis. The array alternates sub-apertures transmitting XLP and YLP at $660\,\mathrm{nm}$, enabling simultaneous, co-registered polarization-resolved views in a single exposure. The optical layout is summarized in Figure~\ref{fig:results2}(b), where collimated $660\,\mathrm{nm}$ illumination passes the specimen (Plane~1), a diffuser and $20\times$ objective form an intermediate image (Plane~2), and the beam is recollimated (Plane~3) before reaching the dual metalens array (Plane~4). The metalens focuses orthogonal linear polarizations into adjacent sub-images, which are re-imaged by a $50\times$ objective (Plane~5) onto the monochrome sensor (Plane~6). This configuration eliminates sequential acquisitions and mechanical rotation, removes inter-frame registration errors, and reduces measurement time while preserving polarization contrast.

\subsubsection{Histogram analysis across leaf cohorts}

To assess chlorophyll content and spatial heterogeneity, pixel intensity histograms were computed from polarization-resolved metalens images (Figure~\ref{fig:results2}(d)). Raw 8-bit grayscale data were recorded at $660,\mathrm{nm}$ for the XLP and YLP channels. Exposure was identical for L1--L3, whereas L4 required a lower exposure setting because of its reduced chlorophyll content and higher transmittance. YLP frames were registered to their XLP counterparts using ECC alignment, and histograms were calculated only within the overlapping region. The normalized distributions revealed a clear separation between XLP and YLP for all leaves, with YLP generally exhibiting a broader range and higher peak values, a trend consistent with polarization-dependent contrast influenced by tissue anisotropy, pigment concentration, and scattering. In L1 and L2, the YLP channel displayed a distinct low-intensity mode, indicating stronger absorption in that polarization state. L4 produced a narrower, right-shifted distribution, a result of both exposure adjustment and reduced chlorophyll absorption.

\subsubsection{Localized analysis}

Since the SRN metalens array operates at 660\,nm with high transmission efficiency, matching the chlorophyll\mbox{-}a absorption band, the modality is intrinsically sensitive to chlorophyll content and therefore to early deficiency. In L3, reduced pigment leads to weaker absorption and higher transmittance, which elevates the recorded intensity in both linear channels and facilitates early detection. Figure~\ref{fig:results2}(e) presents pseudo-colored maps that accentuate this contrast. The polarization-suppressed component $I_{\mathrm{component}}$ (eq. \ref{eq:icomp}) down-weighs pixels with strong polarization contrast and isolates the isotropic, absorption-driven signal, revealing subtle chlorophyll loss that may be missed in a single linear state. Figure~\ref{fig:results2}(f) analyzes two subregions (L1-2 and L1-3) from the same healthy leaf to compare products derived from the co-registered XLP and YLP measurements. The reconstructed \emph{Unpol Avg}—formed from the two polarization channels—suppresses polarization-odd components and uncorrelated noise, producing smoother venous boundaries than either single channel. In both subregions the \emph{Abs Diff} map isolates anisotropic signal associated with vein walls and oriented mesophyll textures, while the \emph{Signed Diff} map preserves the polarity of the contrast (proportional to $S_1/S_0$) and therefore encodes the local analyzer orientation that maximizes transmission. Sobel edge maps emphasize these trends: edges from XLP and YLP highlight complementary orientations, edges from \emph{Abs Diff} concentrate along anisotropic structures, and edges from the \emph{Unpol Avg} exhibit superior continuity because of reduced noise. Together, these products provide higher local contrast-to-noise and more reliable boundary fidelity than any single linear state, and they deliver information—orientation and anisotropy—that a polarization-insensitive capture cannot supply.

\begin{equation}
I_{\mathrm{component}}(x,y)=\big(I_x(x,y)+I_y(x,y)\big)\!\left(1-\frac{\left|I_x(x,y)-I_y(x,y)\right|}{I_x(x,y)+I_y(x,y)+\varepsilon}\right),
\label{eq:icomp}
\end{equation}\smallskip

where $\varepsilon > 0$ prevents division by zero. 

In our measurements, the SRN metalens array showed polarization-dependent image features that matched the type of contrast we had previously observed with the objective-lens setup. The XLP and YLP channels often revealed different aspects of the leaf structure, with some regions appearing brighter in one channel while becoming less visible in the other. Because leaf tissue combines birefringence, scattering, and depolarization, it is difficult to link every feature to a single analyzer orientation, yet the two polarization channels still exposed structural details and contrast variations that a conventional unpolarized system would not capture. Through image processing, the \emph{Abs Diff} and \emph{Signed Diff} maps brought out vein edges, oriented mesophyll textures, and areas with clear contrast reversal. The \emph{Unpol Avg} images, by contrast, reduced polarization-specific contributions and produced smoother outlines. We did notice that the metalens results have a slightly coarser texture than the objective-lens images, which is due to the smaller effective aperture or other differences in the optical path. Nevertheless, edge sharpness and polarization contrast were maintained. Importantly, the integrated metalens array captures both polarization states in a single shot, ensuring inherent co-registration and avoiding motion or alignment errors. With its flat, CMOS-compatible design, the device can be built into compact systems, and we expect further improvement with tighter fabrication control and more stable experimental conditions.

\section{Conclusion}
This work aimed to evaluate a silicon-rich nitride (SRN) metalens array for dual-polarization imaging at the chlorophyll-a absorption peak of 660\,nm. The device integrates polarization selection and focusing into a compact, CMOS-compatible platform, enabling simultaneous acquisition of XLP and YLP channels. Our results show that the metalens produces polarization-resolved features consistent with those from a conventional objective-lens setup, with \emph{Abs Diff} and \emph{Signed Diff} maps revealing vein boundaries, mesophyll textures, and contrast reversals, while \emph{Unpol Avg} images reduced polarization-specific components. The system’s inherent co-registration eliminates alignment errors, offering a practical advantage over sequential methods. A slightly coarser texture was observed in metalens images, due to aperture and optical-path differences, representing a current limitation. The approach is relevant for compact, in-field plant imaging systems, and future work will focus on extending the design to capture additional polarization states, thereby increasing the sensitivity and information content of the measurements.

\bibliography{sn-bibliography}

\end{document}